\documentclass[9pt,twocolumn,twoside]{opticajnl}
\journal{opticajournal} 

\setboolean{shortarticle}{true}

\newcommand{\note}[1]{\textcolor{black}{ #1}}

\newcommand{\crg}[1]{\textcolor{orange}{ #1}}

\usepackage{color}
\usepackage{soul}
\usepackage{lineno}
\title{Non-orthogonal Transformations of Structured Light Using Ellipticity-Dependent Ince–Gaussian Modes}
\author[1]{Dayver Daza-Salgado}
\author[1]{Edgar Medina-Segura}
\author[1,*]{Carmelo Rosales-Guzmán}

\affil[1]{Centro de Investigaciones en Óptica, A.C., Loma del Bosque 115, Lomas del Campestre, 37150,   León, Guanajuato, México}

\affil[*]{carmelorosalesg@cio.mx}

\begin{abstract}
The Ince–Gaussian modes form a complete set of solutions to the paraxial wave equation parametrized by an ellipticity parameter $\varepsilon$, enabling a continuous transition between Laguerre– and Hermite–Gaussian modes. While each fixed $\varepsilon$ defines an orthogonal basis, modes associated with different ellipticities are not mutually orthogonal, and no explicit transformation between such bases has been reported. Here, we derive the first \note{explicit finite analytical expression} to transformation between Ince–Gaussian bases of arbitrary ellipticity, enabling direct and experimentally accessible mapping between non-orthogonal structured-light representations. We further demonstrate an experimental implementation using spatial light modulators to perform ellipticity-resolved modal decomposition. This framework introduces ellipticity as a controllable degree of freedom for structured light engineering, enabling new strategies for mode conversion, encoding, and high-dimensional optical information processing.
\end{abstract}

\setboolean{displaycopyright}{false} 

\begin{document}
\maketitle

The ability to tailor light across its various degrees of freedom—including phase, amplitude, and polarization—has given rise to the concept of structured light fields \cite{Roadmap,Shen2022,forbes2019structured}. This concept has permeated both classical and quantum regimes, enabling a wide range of applications in areas such as optical communications, optical tweezers, and metrology \cite{rosales2024perspective,yang2021,ndagano2017creation}. As a result, the development of new theoretical and experimental tools for controlling the properties of light remains a central topic in modern optics. Structured light fields are commonly described using complete sets of solutions to the paraxial wave equation, most prominently Hermite–Gaussian (HG), Laguerre–Gaussian (LG) \cite{siegman1973hermite}, and Ince–Gaussian (IG) modes \cite{bandres2004ince}. Among these, IG modes constitute the most general separable family, providing a continuous interpolation between Cartesian and circular symmetries through an ellipticity parameter $\varepsilon$, which controls the ellipticity of their transverse intensity and phase profiles and connects the LG limit ($\varepsilon=0$) to the HG limit ($\varepsilon \rightarrow \infty$). For any fixed $\varepsilon$, IG modes form a complete orthogonal basis; however, bases associated with different ellipticities are not mutually orthogonal. Despite this generality, the relationship between IG bases of different ellipticities remains largely unexplored. Existing modal decomposition techniques operate almost exclusively within a single orthogonal basis, and \note{explicit finite analytical expression} across parameter-dependent bases are rarely available. Consequently, ellipticity—although a natural geometric descriptor of structured light and in particular IG modes—has not been fully exploited as a functional degree of freedom. Importantly, even though IG modes can be expressed in terms of LG modes and vice versa \cite{bandres2004ince}, no explicit formulation has been established to directly relate IG modes defined at different ellipticities. In particular, no previous work provides a direct \note{analytical expression for} mapping between IG bases of different ellipticities without relying on intermediate numerical decompositions or basis truncations.

In this Letter, we derive a compact analytic expression that directly connects IG modes of arbitrary ellipticity through finite superpositions. This formulation establishes ellipticity as a continuous transformation parameter between non-orthogonal bases and provides a unified framework linking IG, LG, and HG representations. Unlike conventional unitary mode transformations, this approach enables controlled redistribution of modal weight across non-orthogonal bases, introducing additional flexibility for modal control. We further demonstrate an experimental implementation based on spatial light modulators, performing ellipticity-resolved modal decomposition and validating the predicted inter-basis coefficients \cite{schulze2012wavefront,sroor2021modal}. These results establish a practical framework for exploiting ellipticity as a tunable degree of freedom in wavefront design and in particular structured-light engineering, with potential applications in optical information encoding.

The set of IG$_{p,m}^{\varepsilon,\sigma}$ form a complete orthogonal basis for a fixed $\varepsilon$, indexed by order $p$, degree $m$ and parity $\sigma$. To relate modes of different ellipticities, we employ LG modes as an intermediate basis. \note{The expansion of IG modes into the Laguerre–Gaussian basis can be written as,} \cite{PhysRevA.87.033806, bandres2004ince,abramochkin2025helical}
\begin{equation}
IG_{p,m}^{\sigma,\varepsilon}=\sum_{l}^pD_{l,n}(\varepsilon)\,LG_{n,l}^{\sigma},
\label{eq:3}
\end{equation}
while LG modes admit the inverse relation,
\begin{equation}
LG_{n,l}^{\sigma}=\sum_{m}^pB_{m}^{l,n}(\varepsilon)IG_{p,m}^{\sigma,\varepsilon},
\label{eq:5}
\end{equation}
where their parameter are related by $n=(p-m)/2$ and $l=m$. 

Combining both expressions yields a direct transformation between IG bases of ellipticities $\varepsilon_1$ and $\varepsilon_2$ (see \crg{\bf Supplementary materials} for a detailed derivation), 
\begin{align}
IG_{p,m_1}^{\sigma,\varepsilon_1} =
\sum_{m_2}^p c_{p,m_2}(\varepsilon_1,\varepsilon_2)
IG_{p,m_2}^{\sigma,\varepsilon_2},
\label{eq:6}
\end{align}
which are constrained to the same index $p$ and whose complex weighting coefficients $c_{p,m_2}(\varepsilon_1,\varepsilon_2)$ are given by,
\begin{equation}
c_{p,m_2}(\varepsilon_1,\varepsilon_2)=\sum_{l}^pD_{l,n}(\varepsilon_1)\,B_{m_2}^{l,n}(\varepsilon_2).
\label{eq:7}
\end{equation}

Equivalently, the coefficients $c_{p,m_2}$ are given by overlap integrals,
\begin{equation}
\begin{split}
&c_{p,m_2}=\iint \limits_{-\infty}^{\infty}
IG_{p_1,m_1}^{\sigma_1,\varepsilon_1}\,
(IG_{p_2,m_2}^{\sigma_2,\varepsilon_2})^{*}\,
d\xi d\eta \\ =&
\delta_{\sigma_1\sigma_2}\,\delta_{p_1p_2}\,
(-1)^{p+\frac{m_1+m_2}{2}}\,\sum_l^p(1+\delta_{0,l})\,\Gamma(n+l+1)\,n!
\times\\
&A_{(l+\delta_{o,\sigma_1})/2}^{\sigma_1}(a_p^{m_1},\varepsilon_1)\,
A_{(l+\delta_{o,\sigma_2})/2}^{\sigma_2}(a_p^{m_2},\varepsilon_2).
\label{eq:8}    
\end{split}
\end{equation} 
The terms $A^{\sigma}_{(l+\delta_{0,\sigma})/2}(\varepsilon)$ are the Fourier coefficients of the Ince polynomials, which encode the ellipticity dependence of each mode. Together, these terms give the total overlap between the two IG modes of different ellipticities. 

Equation \ref{eq:8} constitutes the central result of this manuscript, providing \note{an explicit and compact analytical expression that directly relates IG modes of arbitrary ellipticity}. Although this transformation can be formally constructed via intermediate LG expansions, the resulting expression provides the first direct analytic mapping between IG bases of arbitrary ellipticity. Another important aspect of Eq. \ref{eq:8} is that, as expected, it reduces to the orthogonality relation for IG modes when $\varepsilon_1=\varepsilon_2$, \eqref{eq:8} \cite{bandres2004ince},
\begin{equation}
    \iint \limits_{-\infty}^{\infty} IG_{p_1,m_1}^{\sigma_1,\varepsilon}(IG_{p_2,m_2}^{\sigma_2,\varepsilon})^{*}d\xi d\eta=\delta_{\sigma_1\sigma_2}\delta_{p_1p_2}\delta_{m_1m_2},
\end{equation}
which are orthogonal in parity $\sigma$ and indices $p$ and $m$, see \crg{\bf Supplementary materials} for a detailed derivation. In contrast, when $\varepsilon_1 \neq \varepsilon_2$, orthogonality is lost in the modal index $m$ but preserved for parity $\sigma$ and index $p$. Finally, the number of base modes of IG$^{\varepsilon_2}$ can be determined from $p$ by considering the allowed pairs $(p,m)$ under the parity conditions of the IG modes.
Specifically, if $p$ is even, the number of basis modes is $N_p^e=(p+2\delta_{\sigma,e})/2$, while for odd $p$ the number is $N_p^o=(p+1)/2$ \cite{bandres2004ince}. 

Importantly, Eq. \ref{eq:8} reveals that IG modes with different ellipticities form a family of mutually non-orthogonal but complete bases, analogous to overcomplete representations in other areas of physics \cite{Overcomplete}. In this context, the ellipticity $\varepsilon$ acts as a continuous parameter that deforms the modal basis, with the overlap coefficients quantifying the "distance'' between bases in Hilbert space. This interpretation provides a geometric perspective on structured light, where transformations between coordinate systems correspond to controlled redistributions of modal weight. Such non-orthogonal basis transformations are fundamentally distinct from conventional unitary mode conversions and may offer new flexibility in modal analysis and encoding. 

\begin{figure}[tb]
    \centering \includegraphics[width=0.95\linewidth]{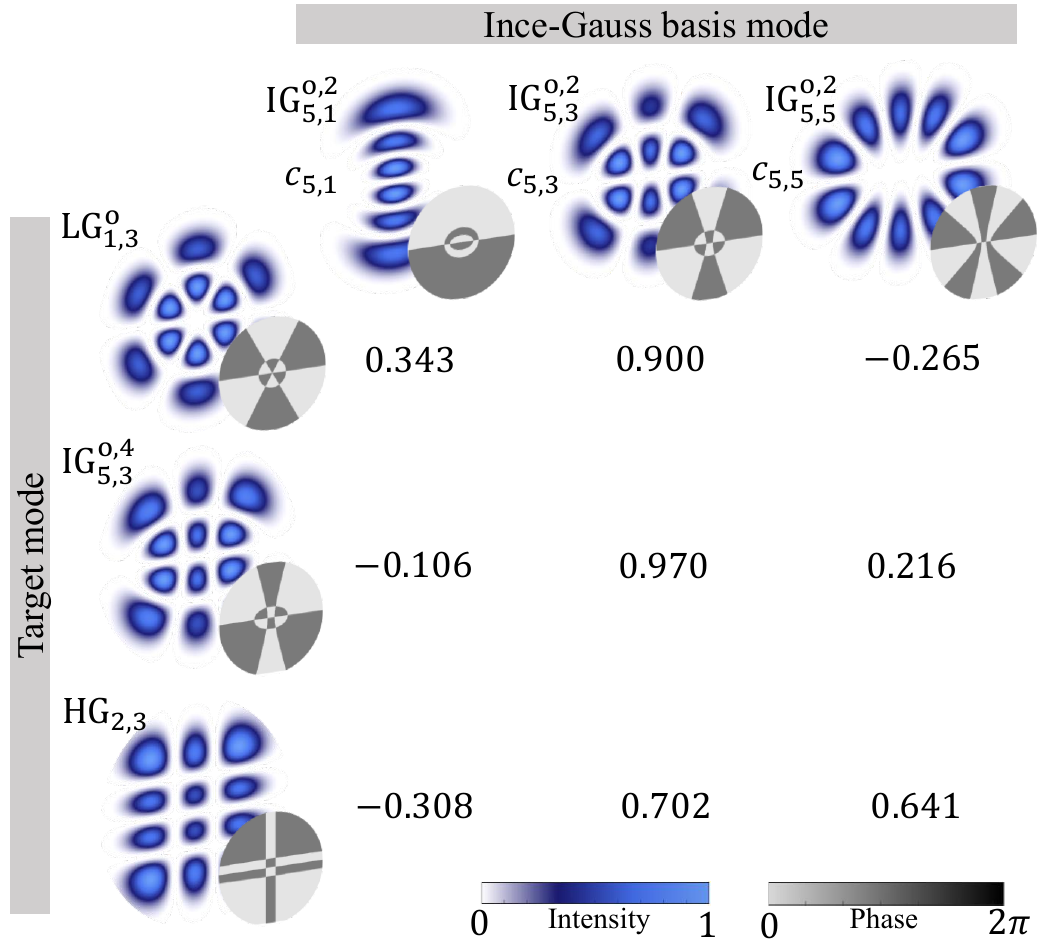}
    \caption{Decomposition of representative structured-light modes into an Ince–Gaussian (IG) basis of fixed ellipticity ($\varepsilon_2=2$). Top row: basis modes IG$_{5,m_2}^{o,\varepsilon_2}$ with $m_2=1,3,5$. Bottom rows: target modes (Laguerre–Gaussian, Ince–Gaussian, and Hermite–Gaussian) expressed as finite superpositions of the basis elements. The complex coefficients $c_{5,m}$ obtained from Eq. \ref{eq:8} quantify the contribution of each basis mode. The results illustrate the redistribution of modal weight arising from ellipticity mismatch, evidencing the non-orthogonality between IG bases of different $\varepsilon_2$}
    \label{ExpandedModes}
\end{figure}

Figure \ref{ExpandedModes} illustrates the decomposition of representative LG, IG, and HG modes into an IG basis of fixed ellipticity. For a given order $p$, only a finite number of modes contribute, consistent with the derived expressions. More precisely, here we show the modal decomposition of the modes LG$_{1,3}^o$, IG$_{5,3}^{o,4}$ and HG$_{2,3}$ in the IG basis of $\varepsilon_2 = 2$. Since $p=5$, there are only 3 constituting IG modes, namely $\mathrm{IG}_{5,1}^{o,2}$, $ \mathrm{IG}_{5,3}^{o,2}$ and $ \mathrm{IG}_{5,5}^{o,2}$. In the figure the transverse intensity profile and phase distribution of modes LG, IG and HG are represented from top to bottom and calculated at the plane $z=0$. In a similar way, the intensity and phase profile of the constituting IG modes is shown on the top. The complex coefficients $c_{p,m_2}$ corresponding to each IG mode, computed from Eq. \ref{eq:8} and labelled $c_{5,1}$, $c_{5,3}$ and $c_{5,5}$, are shown in the rows for the LG, IG and HG modes. Notice that for this example we used for the HG mode  $\varepsilon_1=1000$.

\begin{figure}[tb]
    \centering \includegraphics[width=1\linewidth]{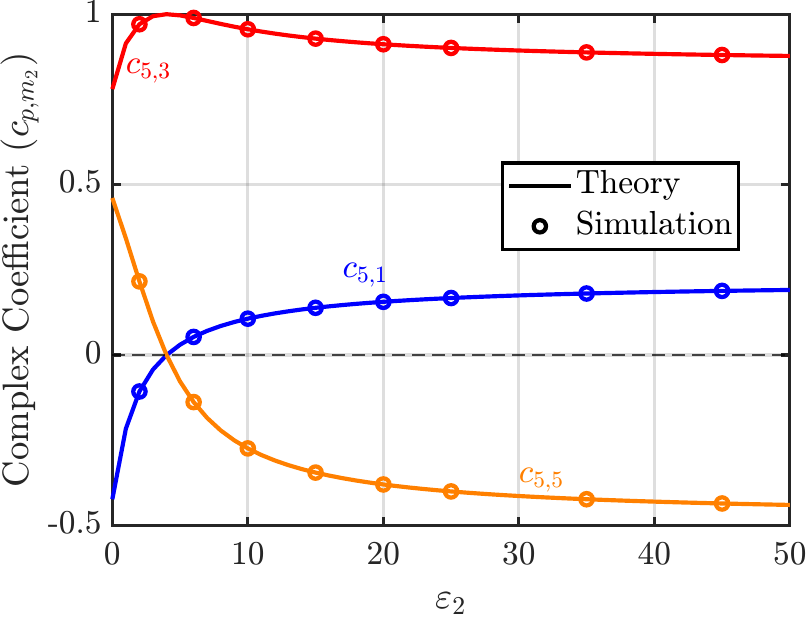}
    \caption{Ellipticity-dependent modal coefficients $c_{p,m}$ describing the overlap between a fixed input mode IG$_{5,3}^{o,\varepsilon_1}$ and IG basis modes IG$_{5,m}^{o,\varepsilon_2}$ as a function of $\varepsilon_2$. Solid lines: analytic prediction from \eqref{eq:6}. Symbols: numerical simulations. The peak at $\varepsilon_1=\varepsilon_2$ indicates recovery of orthogonality, while deviations from this point lead to continuous redistribution of modal weight across modes. This behavior demonstrates the deformation of the modal basis and the emergence of non-orthogonal representations controlled by ellipticity.}
    \label{Coef}
\end{figure}

Additionally, a plot of the modal weight redistribution as a function of $\varepsilon_2$, for values in the interval $[0,50]$ is shown in Fig. \ref{Coef} for the IG$_{5,3}^{o,4}$ mode and for the three complex coefficients $c_{5,1}$, $c_{5,3}$ and $c_{5,5}$. The solid line represent the exact values obtained from \eqref{eq:8}, whereas the dots represent values from numerical simulations obtained through modal projections. As expected, an increase in $\varepsilon_2$ leads to a redistribution of the modal weights. In particular, when $\varepsilon_1\neq\varepsilon_2$, the complex coefficients associated with different indices $(p,m)=(5,1)$ and $(5,5)$, which vanish for identical ellipticities, acquire non-zero values. Conversely, the coefficient corresponding to the mode with identical indices $(5,3)$ reaches its maximum when $\varepsilon_1=\varepsilon_2$, and gradually decreases as the ellipticity mismatch increases.

The theoretical framework described above can be implemented experimentally using direct measurement of inter-basis overlap coefficients \cite{schulze2012wavefront,Forbes2016,pinnell2020modal,ZhaoBo2026,SPIEbook}. In this approach, each of the complex modal coefficient $c_{p,m_2}=W_{p,m_2}\exp{(i\alpha_{p,m_2})}$, is obtained through an optical inner product between the target mode and the elements of the basis. Experimentally, the inner product is implemented by projecting the target field into a series of suitable computer generate holograms (CGH) displayed on a Spatial Light Modulator (SLM), whereby $W_{p,m_2}$ is measured as the on-axis intensity in the far field, that is $|W_{p,m_2}|^2=I_{p,m_2}(0,0)$, see \cite{SPIEbook} for details. In addition, the intermodal phase $\alpha_{p,m_2}$ is determined in a similar way using the relation (see \cite{Forbes2016} for more details),
\begin{equation}
\alpha_{p,m_2}=\arctan\!\left[\frac{2I_{p,m_2}^{\sin}-I_{p,m_2}^{0}-I_{p,m_2}}{2I_{p,m_2}^{\cos}-I_{p,m_2}^{0}-I_{p,m_2}}\right],
\label{eq:12}
\end{equation}
where $I^0_{p,m_2}$ and $I_{p,m_2}$ are the on-axis intensity of the projection with a reference mode and the rest of the modes, respectively. Moreover, $I_{p,m_2}^{\sin}$ and $I_{p,m_2}^{\cos}$ represent the on-axis intensity of the projection with two auxiliary superposition modes given by,
\begin{equation}
    \begin{split}
        T^{\sin}=\text{IG}_{p,m_2}^{\varepsilon_2,0}+i\text{IG}_{p,m_2}^{\varepsilon_2}\\
        T^{\cos}=\text{IG}_{p,m_2}^{\varepsilon_2,0}+\text{IG}_{p,m_2}^{\varepsilon_2}.
    \end{split}
\end{equation}

\begin{figure}[tb]
    \centering \includegraphics[width=1\linewidth]{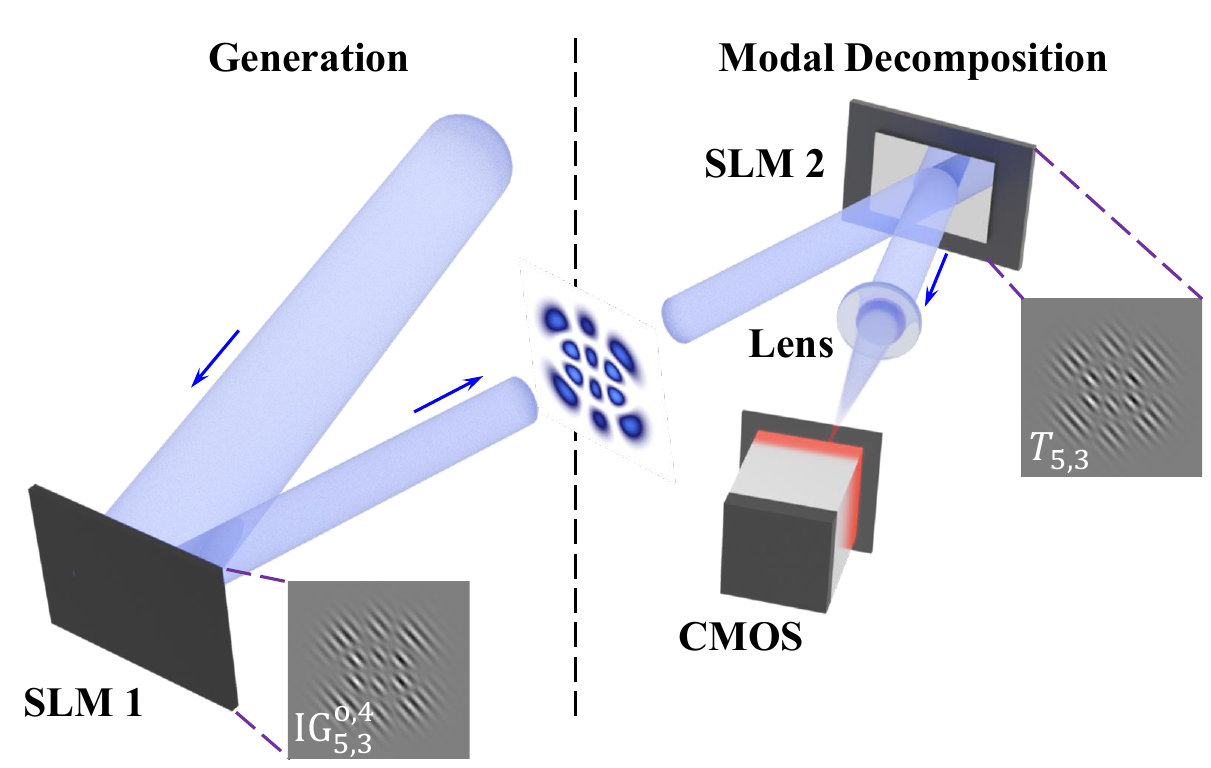}
    \caption{Experimental implementation of ellipticity-resolved modal decomposition. A target IG mode is generated using complex-amplitude modulation on SLM 1 and projected onto IG basis modes of different ellipticity using SLM 2. The overlap coefficients are obtained from on-axis intensities in the Fourier plane. This setup enables direct measurement of inter-basis coefficients $c_{p.m_2}$ and experimental validation of the ellipticity-dependent transformation.}
    \label{Setup}
\end{figure}

\begin{figure}[tb]
    \centering \includegraphics[width=1\linewidth]{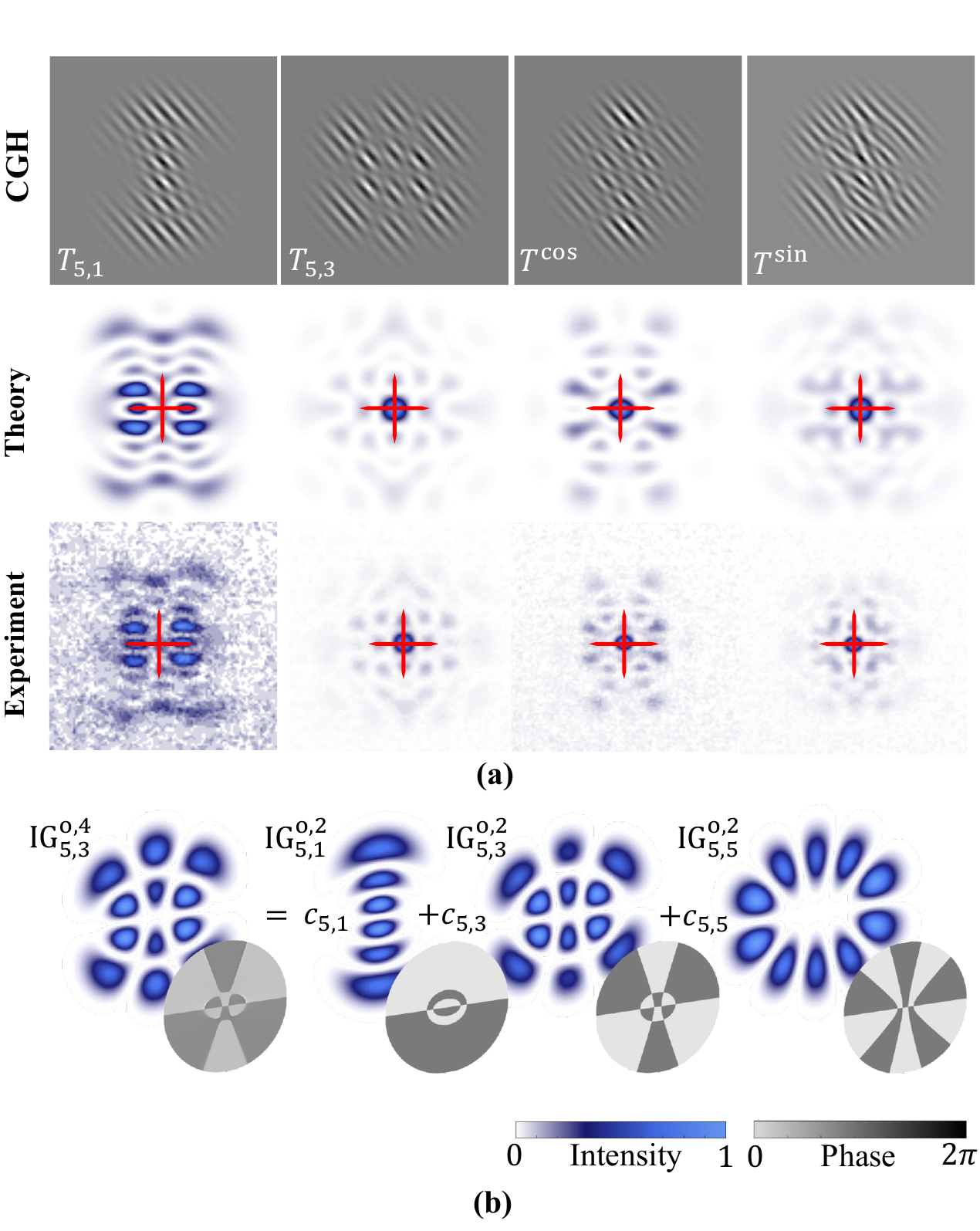}
    \caption{Experimental measurement and reconstruction of modal coefficients. (a) Representative computer-generated holograms (top), simulated (middle), and measured (bottom) far-field intensity distributions used to extract modal amplitudes and phases; the red marker indicates the on-axis measurement point. (b) Reconstruction of the input mode from measured coefficients, showing intensity and phase. The agreement with theory confirms accurate retrieval of complex coefficients and validates the non-orthogonal modal expansion.}
    \label{CGH}
\end{figure}

To demonstrate experimentally how a given IG$^{\varepsilon_1}$ is decomposed as the superposition of IG$^{\varepsilon_2}$ using the technique described above, we implemented the experimental setup depicted in Fig.~\ref{Setup}. Here, the IG$^{\varepsilon_1}$ mode are first generated using complex-amplitude modulation \cite{arrizon2007pixelated} implemented on a liquid-crystal SLM (Holoeye Pluto 2.1, $1920\times1080$ pixels, $8~\mu$m pitch). To this end, a linearly polarized He-Ne laser ($\lambda=633$~nm) is expanded and collimated before illuminating {\bf SLM 1}, see Fig. \ref{Setup} {\bf Generation} section. \note{The amplitudes retrieved from the on-axis intensity measurements are initially in arbitrary detector units. To compare the results with theory and reconstruct the field, the complex coefficients were normalized so that $\sum_m|c_{p.m}|^2=1$, following standard modal-decomposition procedures.} By way of example, the bottom-left inset of Fig. \ref{Setup}(a) shows the required hologram to generate the IG$_{5,3}^{o,4}$ mode. It is worth mentioning that the first diffraction order, which contains the desired mode, was isolated using a spatial filter (not shown here). The generated field is then projected onto {\bf SLM 2} for its modal decomposition, see Fig \ref{Setup} {\bf Modal Decomposition} section. Here the corresponding digital holograms described above to measure the complex coefficients $c_{p,m_2}$ (amplitude and phase) are sequentially displayed one by one on the SLM until all the required intensity values are computed. The resulting field is then Fourier transformed using a lens $L$ of focal length ($f=250$ mm) in a $2f$ configuration, producing the spatial-frequency spectrum at its back focal plane. Here, the on-axis intensity of each projection, is recorded with a High-Sensitive USB 3.0 CMOS camera (Thorlabs, $1280\times1024$ pixels). By way of example, typical holograms to perform the modal projection are shown in Fig. \ref{CGH}(a), along with the corresponding far-field intensity distribution for each projection. Here, the red cross marks the position where the on-axis intensity is measured in each case. We also show numerical simulation of the far-field intensity distribution to highlight the accuracy of the experimental technique. Fig. \ref{CGH}(b) illustrates a representative case showing the reconstructed intensity and phase of the IG$_{5,3}^{o,4}$ mode using experimental measurements in the $\{{\mathrm{IG}_{5,1}^{o,2}, \mathrm{IG}_{5,3}^{o,2}, \mathrm{IG}_{5,5}^{o,2}}\}$ base according to \eqref{eq:6}. The resulting reconstruction confirms the recovery of amplitude and phase information from the experimental coefficients, which are given by $c_{5,1} = 0.1124\exp(-i137.8129^{\circ})$, $c_{5,3} = 0.9659\exp(-i159.3663^{\circ})$ and $c_{5,5} = 0.2333\exp(i161.6038^{\circ})$. It is worth emphasizing that the experimental retrieval of complex modal coefficients requires careful alignment and calibration of the optical system. Small inaccuracies in the optical overlap, detector response, or CGH encoding can propagate into errors in the measured modal amplitudes and intermodal phases. In this experiment, an imperfect calibration of the camera-measured energy led to deviations in the retrieved phase values; however, the reconstructed mode remains in good agreement with the theoretical field. \note{It is worth emphasising that The theoretical curves in Fig. \ref{Coef} represent the real-valued coefficients as a function of ellipticity. In contrast, the experimentally retrieved coefficients are complex, and the sign of the real theoretical coefficient is encoded in the phase of the complex coefficient (0 or $\pi$ for this mode family). The measured phases of $c_{5,1}$, $c_{5,3}$ and $c_{5,5}$ cluster around $\pm180\deg$, consistent with the sign structure predicted in Fig. \ref{Coef}}

In summary, we have \note{presented an explicit finite analytical expression that allows} to transform between Ince–Gaussian (IG) modes of arbitrary ellipticity, showing that modes defined in one elliptic basis can be expressed as a finite superposition of modes from another. This transformation emerges from analytic overlap coefficients governed by the Fourier components of the Ince polynomials and preserves orthogonality only in the modal order and parity. The resulting framework treats ellipticity as a continuous parameter that deforms the IG basis, providing controlled access to non-orthogonal structured light representations. We further implemented an experimental scheme based on spatial light modulators to measure these inter-basis coefficients directly, demonstrating close agreement with the theoretical predictions. This confirms that the modal relationships derived here are not only analytically exact but also experimentally accessible. By enabling explicit and measurable transformations between ellipticity-dependent IG bases, this work introduces a practical approach to ellipticity-resolved modal analysis and expands the toolbox for structured light engineering beyond conventional orthogonal modal frameworks. This capability enables adaptive basis selection in modal encoding schemes, where ellipticity can be tuned to optimize information capacity, minimize modal cross-talk, or match system-specific propagation conditions. In contrast to fixed orthogonal bases, this approach provides additional flexibility for tailoring modal representations to physical constraints.
\begin{backmatter}
\bmsection{Funding} 
D.D.S. (CVU: 1159764) and E.M.S. (CVU: 742790) acknowledge financial support from SECIHTI-México through PhD scholarships, C.R.G. through the project CBF-2025-I-1804.

\bmsection{Disclosures} 
The authors declare no conflicts of interest.

\bmsection{Supplemental document} See Supplement 1 for supporting content

\section{References}
\end{backmatter}

\end{document}